\documentclass[conference,a4paper]{IEEEtran}
\IEEEoverridecommandlockouts

\usepackage[dvips]{graphicx}
\usepackage{subfigure}
\usepackage{eclbkbox}
\usepackage{indent}

\makeatletter
\def\tbcaption{\def\@captype{table}\caption}
\def\figcaption{\def\@captype{figure}\caption}

\ifCLASSINFOpdf
\else
\fi

\begin{document}

\title{Tourist Navigation in Android Smartphone\\by using Emotion Generating Calculations and Mental State Transition Networks
\thanks{\copyright 2012 IEEE. Personal use of this material is permitted. Permission from IEEE must be obtained for all other uses, in any current or future media, including reprinting/republishing this material for advertising or promotional purposes, creating new collective works, for resale or redistribution to servers or lists, or reuse of any copyrighted component of this work in other works.}
}

\author{\IEEEauthorblockN{Takumi Ichimura}
\IEEEauthorblockA{Faculty of Management and\\ Information Systems,\\
Prefectural University of Hiroshima\\
1-1-71, Ujina-Higashi, Minami-ku,\\
Hiroshima, 734-8558, Japan\\
Email: ichimura@pu-hiroshima.ac.jp}

\and
\IEEEauthorblockN{Kosuke Tanabe}
\IEEEauthorblockA{Graduate School of Comprehensive\\ Scientific Research\\
Prefectural University of Hiroshima,\\
1-1-71, Ujina-Higashi, Minami-ku,\\
Hiroshima, 734-8558, Japan\\
Email: bakabonn009@gmail.com}

\and
\IEEEauthorblockN{Issei Tachibana}
\IEEEauthorblockA{Faculty of Management\\ and Information Systems,\\
Prefectural University of Hiroshima\\
1-1-71, Ujina-Higashi, Minami-ku,\\
Hiroshima, 734-8558, Japan\\
Email: isseing1224@gmail.com}
}

\maketitle

\begin{abstract}
Mental State Transition Network which consists of mental states connected to each other is a basic concept of approximating to human psychological and mental responses. It can represent transition from an emotional state to other one with stimulus by calculating Emotion Generating Calculations method. A computer agent can transit a mental state in MSTN based on analysis of emotion by EGC method. In this paper, the Andorid EGC which the agent works in Android smartphone can evaluate the feelings in the conversation. The tourist navigation system with the proposed technique in this paper will be expected to be an emotional oriented interface in Android smartphone.
\end{abstract}

\IEEEpeerreviewmaketitle

\section{Introduction}
\label{sec:Introduction}
At a beginning of the 21th century, some people said that although the 20th century was the ``time of matter,'' the 21st century will be the ``time of the heart.'' The expression ``time of the heart'' indicates mental health care, change of the sense of values considering emotion, and so on. Researchers in various fields such as information science, psychology \cite{Fujiwara10,Strongman96,Power07}, human engineering \cite{Kimura08,Nagamachi10}, brain physiology \cite{Matsunaga09,Foster08} and so on are approaching to `mind.' It was difficult to deal with mind and emotion in science field because they are ambiguous and vague to formulate a way to measure the degree of them. However, recent researches can approach the process of the mind scientifically by development of measurement machines, measurement method using computer which simulates the mind\cite{Park10}. Emotions such as love, hate, courage, fear, joy, sadness, pleasure, and disgust are represented in both psychological and physiological terms. An essential role in working of the mind was analyzed in philosophy, psychology, and from the learning and cognition perspective.

Our research group proposed an estimation method to calculate the agent's emotion from the contents of utterances and to express emotions which are aroused in computer agent by using synthesized facial expression \cite{Ichimura03, Mera03a, Mera03b, Mera05}. Emotion Generating Calculations (EGC) method \cite{Mera03a} based on the Emotion Eliciting Condition Theory \cite{Elliott92} can decide whether an event arouses pleasure or not and quantify the degree under the event.

Calculated emotions effect to the mood of the agent. Ren \cite{Ren06} describes Mental State Transition Network (MSTN) which is the basic concept of approximating to human psychological and mental responses. The assumption of discrete emotion state is that human emotion is classified into some kinds of stable discrete states, called ``mental state,'' and the variance of emotions occurs in the transition from a state to other state with an arbitrary probability. Mera \cite{Mera10b} developed a computer agent that can transit a mental state in MSTN based on analysis of emotion by EGC method. EGC calculates emotion and the type of the aroused emotion is used to transit mental state \cite{Mera10b}.

Recently, mobile phones have infiltrated in our lives and have made their own unique stand. Android smartphone has been widely used in various ways and would be also replacing the laptop by enabling internet access through the smartphone. The users can use not only phone feature but also their favorite applications through the application market. Especially, the smartphone user can not only obtain the variety of information but also converse with the agent in a smartphone, because the interface between human and smartphone has been equipped with the speech recognition. Moreover, smartphones have the GPS device and acceleration sensor. They have also popularized the camera feature. Smartphones offer expandable memory features. These devices develop the smartphone potential abilities. Therefore, many applications for smartphones are developed continuously. 

In this paper, we developed Android EGC which the agent works to evaluate the feelings in the conversation in Android smartphone. The proposed technique can be expected to be an emotional orientated interface. The tourist navigation system has been developed as one of the trials for the emotional orientated interface. The system such as the concierge system, Siri or Shabette-Concier, can provide either personal assistance in travel or  associated with information service. In addition, the system can evaluate the feelings of the user at each sightseeing spot by EGC and recommend the restaurant or the shop due to the user current feelings to compare with impression to sightseeing spots. For example, when the user feels ``disgust'', the system guide the spot where the typical user feels ``happy.''

The remainder of this paper is organized as follows. In the section \ref{sec:EGC}, the brief explanation to understand the EGC is described. Section \ref{sec:MentalStateTransitionLarningNetwork} explains to MSTN to measure the current mental state by the stimulus of the EGC. The concierge system which can operate in Android smartphone is proposed in Section \ref{sec:AndroidConciergeSystem}.  In Section \ref{sec:Conclusivediscussion}, we give some discussions to conclude this paper.

\section{Emotion Generating Calculations}
\label{sec:EGC}
\subsection{An Overview of Emotion Generating Process}

\begin{figure}[ht]
\begin{center}
\includegraphics[scale=0.6]{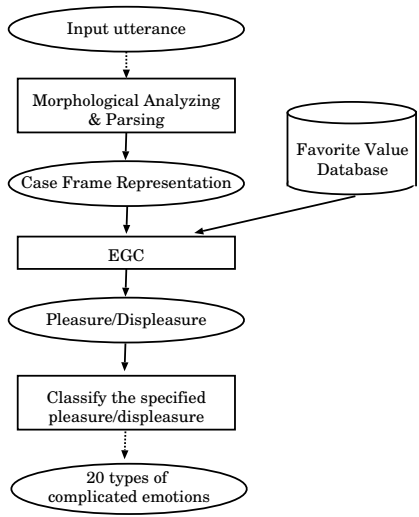}
\caption{Process for generating emotions}
\label{fig:processgeneratingemotion}
\end{center}
\end{figure}

Fig.\ref{fig:processgeneratingemotion} shows the emotion generating process where the user's utterance is transcribed into a case frame representation based on the results of morphological analysis and parsing. The agent works to determine the degree of pleasure/displeasure from the event in case frame representation by using EGC. In the psychological field, ``unpleasure'' is often used as the opposite of ``pleasure.'' However, we use ``displeasure,'' because an explicit intention about ``unhappy'' should be required in case of complex human feelings. EGC consists of 2 or 3 terms such as  subject, object and predicate, which have Favorite Value ($FV$), the strength of the feelings described in section \ref{sec:FavoriteValue}.

Then, the agent divides this simple emotion (pleasure/displeasure) into 20 various emotions based on the Elliott's ``Emotion Eliciting Condition Theory\cite{Elliott92}.'' Elliott's theory requires judging conditions such as ``feeling for another,'' ``prospect and confirmation,'' ``approval/disapproval.'' The detail of this classification method is described in the section \ref{sec:ComplicatedEmotion}.

\subsection{Case Frame Representation}
In order to transcribe the user's utterances into the case frame representation, we implement morphological analysis and parsing to the input sentence. In case of Japanese, JUMAN as a morphological analyzer and KNP as a parser are often used to be embedded system because of their excellent performance and open source \cite{Nasukawa00}.

\subsection{Favorite Value Database}
\label{sec:FavoriteValue}
We calculate pleasure/displeasure about an event by using $FV$. We give positive numbers to some objects when the user likes them, and give negative numbers to other objects which the user dislikes. $FV$ is predefined a real number in the range $[-1.0, 1.0]$.
There are two types of $FV$s, personal $FV$ and default $FV$. Personal $FV$ is stored in a personal database for each person who the agent knows well, and it shows the degree of like/dislike to an object from the person's viewpoint. On the other hand, default $FV$ shows the common degree of like/dislike to an object that the agent feels. Generally, it is generated based on the agent's own preference information according to the result of some questionnaires. Both personal and initial $FV$s are stored in the user own database. An initial value of $FV$ is determined beforehand on the basis of `corpus' of its applied field such as medical informatics. The $FV$s of the objects are gained from a questionnaire. However, there are countless objects in the world. In this paper, we limit the objects that have default $FV$ into the frequently appeared words in the dialogue.

\subsection{Equation of EGC}
We assume an emotional space as three-dimensional space. Therefore, we present a method to distinguish pleasure/displeasure from an event by judging where the ``synthetic vector'' exists \cite{Mera02}.

\begin{table}[tbp]
\begin{center}
\caption{Correspondence between the event type and the axis}
\begin{tabular}{c|c|c|c}
\hline
Event type  & $f_{1}$ & $f_{2}$ & $f_{3}$ \\ \hline 
$V(S)$      &         &        &         \\
$A(S,C)$    &         &        &         \\
$A(S,OF,C)$ &         &        &         \\
$A(S,OT,C)$ & $f_{S}$  &        & $f_{P}$ \\
$A(S,OM,C)$ &         &        &         \\
$A(S,OS,C)$ &         &        &         \\ \hline
$V(S,OF)$   & $f_{S}$ & $f_{OT}-f_{OF}$ & $f_{P}$  \\
$V(S,OT)$   &         &         &         \\ \hline
$V(S,OM)$   & $f_{S}$ & $f_{OM}$ & $f_{P}$  \\ \hline
$V(S,OS)$   & $f_{S}-f_{OS}$ &  & $f_{P}$  \\ \hline
$V(S,O)$    & $f_{S}$ & $f_{O}$ & $f_{P}$  \\ 
            & $f_{O}$ &         & $f_{P}$ \\ \hline
$V(S,O,OF)$ & $f_{O}$ & $f_{OT}-f_{OF}$ & $f_{P}$ \\
$V(S,O,OT)$ &         & $f_{OM}$       &  \\ \hline
$V(S,O,OM)$ & $f_{O}$ & $f_{OM}$ & $f_{O}$ \\ \hline
$V(S,O,O)$ & $f_{O}$ & $f_{I}$  & $f_{P}$ \\ \hline
$V(S,O,OC)$ & $f_{O}$ &          & $f_{OC}$ \\ \hline
$A(S,O,C)$  & $f_{O}$ &          & $f_{P}$ \\ \hline
\end{tabular}
\label{tab:EGC-eventtype}
\end{center}
\end{table}

Table \ref{tab:EGC-eventtype} shows the correspondence between the case element in EGC equations and the axis in the three-dimensional model. In this table, `V(S,*)' is the type of event (verb) and `A(S,*)' is the type of attribute (adjective). 'In this table, each variable is expressed as follows.

\begin{itemize}
\item $f_{S}$ : $FV$ of Subject
\item $f_{OF}$ : $FV$ of Object-From
\item $f_{OM}$ : $FV$ of Object-Mutual
\item $f_{OC}$ : $FV$ of Object-Content
\item $f_{O}$ : $FV$ of Object
\item $f_{OT}$ : $FV$ of Object-To
\item $f_{OS}$ : $FV$ of Object-Source
\item $f_{P}$ : $FV$ of Predicate
\item $f_{I}$ : $FV$ of Instrument or tool
\end{itemize}

\begin{figure}[btp]
\begin{center}
\includegraphics[scale=0.4]{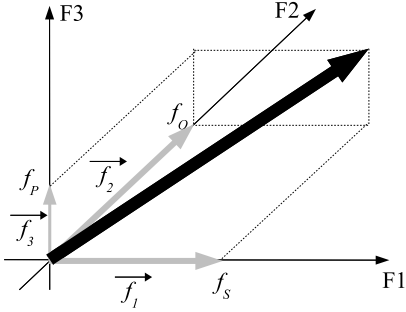}
\caption{Emotion Space for EGC}
\label{fig:EGC-emotionvector}
\end{center}
\end{figure}

Table \ref{tab:EGC-axis}  is a list between the sign of each axis and the generated pleasure/displeasure. When the vector is on the axis, the event does not raise any emotion. When we calculate the synthetic vectors of the events which do not have $f_{2}$ elements, we supply a dummy $FV$, $\beta$ as $f_{2}$ element. We tentatively defined $\beta$ as $+0.5$. Fig.\ref{fig:EGC-emotionvector} is an example of emotion space of event type $V(S, O)$. There are three elements, Subject, Object, and Predicate, in the event type, and the orthogonal vectors by the elements construct a rectangular solid. 

\begin{table}[tbp]
\begin{center}
\caption{pleasure/displeasure in emotional space}
\begin{tabular}{c|c|c|c|c}
\hline
Area & $f_{1}$ & $f_{2}$ & $f_{3}$ & Emotion \\ \hline
I & + & + & + & Pleasure \\
I\hspace{-.1em}I & - & + & + & Displeasure \\
I\hspace{-.1em}I\hspace{-.1em}I & - & - & + & Pleasure \\
I\hspace{-.1em}V & + & - & + & Displeasure \\
V & + & + & - & Displeasure \\
V\hspace{-.1em}I & - & + & - & Pleasure \\
V\hspace{-.1em}I\hspace{-.1em}I & - & - & - & Displeasure \\
V\hspace{-.1em}I\hspace{-.1em}I\hspace{-.1em}I & + & - & - & Pleasure \\ \hline
\end{tabular}
\label{tab:EGC-axis}
\end{center}
\end{table}

\subsection{Complicated Emotion Eliciting Method}
\label{sec:ComplicatedEmotion}
\begin{figure}[btp]
\begin{center}
\includegraphics[scale=0.8]{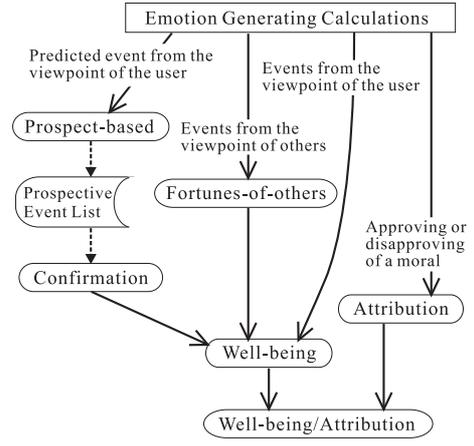}
\caption{Dependency among emotion groups}
\label{fig:EGC}
\end{center}
\end{figure}

Based on emotion values calculated by EGC method and their situations, the pleasure/displeasure is classified into 20 types of emotion. We consider only 20 emotion types, which are classified into an emotional group as follows, ``joy'' and ``distress'' as a group of ``Well-Being,'' ``happy-for,'' ``gloating,'' ``resentment,'' and ``sorry-for'' as a group of ``Fortunes-of-Others,'' ``hope'' and ``fear'' as a group of ``Prospect-based,'' ``satisfaction,'' ``relief,'' ``fears-confirmed,'' and ``disappointment'' as a group of ``Confirmation,'' ``pride,'' ``admiration,'' ``shame,'' and ``disliking'' as a group of ``Attribution,'' ``gratitude,'' ``anger,'' ``gratification,'' and ``remorse'' as a group of ``Well-Being/Attribution'' \cite{Ortony88, Mera03a, Mera03b}. Fig.\ref{fig:EGC} shows the dependency among the groups of emotion types. 

\section{Mental State Transition Learning Network}
\label{sec:MentalStateTransitionLarningNetwork}
\subsection{Mental State Transition Network} 
\begin{figure}[btp]
\begin{center}
\includegraphics[scale=0.5]{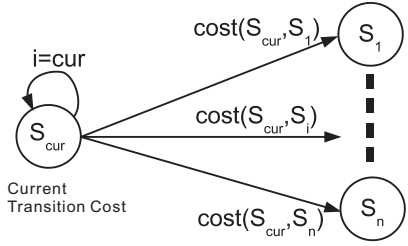}
\caption{Transition Cost}
\label{fig:TransitionCost}
\end{center}
\end{figure}

\begin{figure}[btp]
\begin{center}
\includegraphics[scale=0.6]{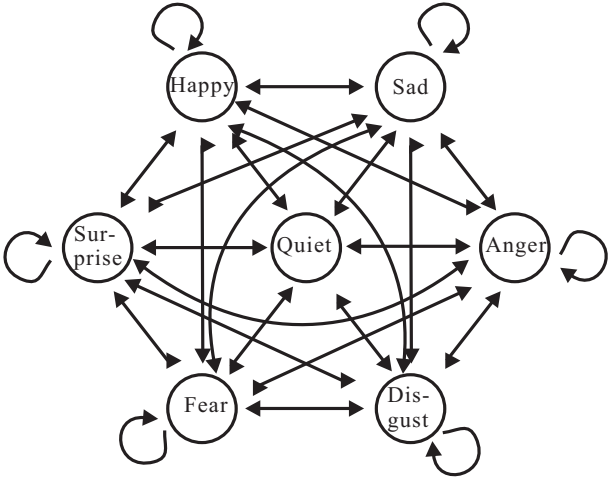}
\caption{Concept of MSTN}
\label{fig:MSTN}
\vspace{-5mm}
\end{center}
\end{figure}

MSTN, proposed by Ren\cite{Ren06}, represents the basic concept of approximating to human physiological and mental responses. He focuses not only information included in the elements of phonation, facial expressions, and speech, but also human psychological characteristics based on the latest achievements of brain science and psychology in order to derive transition networks for human psychological states. The assumption of discrete emotion state is that human emotions are classified into some kinds of stable discrete states, called ``mental state'', and the variance of emotions occurs in the transition from a state to other state with a probability. The probability of transition is called ``transition cost'' and it does not have the same one. Moreover, with no stimulus from the external world, the probability may converge to fall into a certain value as if the confusion of the mind leaves and is relieved. On the contrary, with a stimulus from external world and/or attractive thought in internal world, the continuous accumulated emotional energy cannot jump to the next mental state and remains in its mental state still. The simulated model of mental state transition network\cite{Ren06} describes the simple relations among some kinds of stable emotions and the corresponding transition probability. The probability was calculated from analysis of many statistical questionnaire data.
As shown in Fig.\ref{fig:MSTN}, the MSTN denotes a mental state as a node, a set of some kinds of mental state $\mathcal{S}$, the current emotional state $\mathcal{S}_{cur}$ , and the transition cost $cost(\mathcal{S}_{cur}, \mathcal{S}_{i})$, which is the transition cost as shown in Fig.\ref{fig:TransitionCost}.

In \cite{Ren06}, 6 kinds of mental states and quiet state are considered for questionnaire. That is, the transition table of $cost(\mathcal{S}_{i}, \mathcal{S}_{j})$, $i=1,2,\cdots,7$, $j=1,2,\cdots,7$ is prepared. The experiment for participants was examined without stimulus from external world. Each participant fills in the numerical value from 1 to 10 that means the strength of relation among mental states. Moreover, the same questionnaire was examined under the condition with the stimulus from external world. The 200 participants answered the questionnaire. The numerical values in Table \ref{tab:TransitionCost} show the statistical analysis results. The transition cost from each current state to the next state is summarized to $1.0$.

\begin{table*}[htbp]
\begin{center}
\caption{Transition Cost in MSTN}
\begin{tabular}{l|c|ccccccc}
\hline
\multicolumn{2}{c|}{ }      & \multicolumn{7}{c}{next mental state}\\
\cline{3-9}
\multicolumn{2}{c|}{ }      & happy & quiet & sad   & surprise & angry & fear & disgust\\ \hline
&happy    & 0.421 & 0.362 & 0.061 & 0.060 & 0.027 & 0.034 & 0.032\\
&quiet    & 0.213 & 0.509 & 0.090 & 0.055 & 0.039 & 0.051 & 0.042\\
current&sad      & 0.084 & 0.296 & 0.320 & 0.058 & 0.108 & 0.064 & 0.068\\
mental &surprise & 0.190 & 0.264 & 0.091 & 0.243 & 0.086 & 0.076 & 0.048\\
state  &angry    & 0.056 & 0.262 & 0.123 & 0.075 & 0.293 & 0.069 & 0.121\\
&fear     & 0.050 & 0.244 & 0.137 & 0.101 & 0.096 & 0.279 & 0.092\\
&disgust  & 0.047 & 0.252 & 0.092 & 0.056 & 0.164 & 0.075 & 0.313\\ \hline
\end{tabular}
\label{tab:TransitionCost}
\end{center}
\end{table*}

\subsection{EGC in MSTN}
Even if there is not any signal from external world, the mental state will change small. In this case, the transition costs represented in Table \ref{tab:TransitionCost} are adopted to calculate by using EGC.
In this paper, we assume that the stimulus from external world is the utterance of the user and the transition cost is calculated as follows.

\begin{equation}
cost(\mathcal{S}_{i}, \mathcal{S}_{j})=1-\frac{\#(\mathcal{S}_{i} \rightarrow \mathcal{S}_{j})}{\sum_{j=1}^{7} \#(\mathcal{S}_{i} \rightarrow \mathcal{S}_{j})},
\label{eq:TransitionCost}
\end{equation}
where $\#(\mathcal{S}_{i} \rightarrow \mathcal{S}_{j})$ is the number of transition from mental state $\mathcal{S}_{i}$, $1 \leq i \leq 7$ to $\mathcal{S}_{j}$, $1 \leq j \leq 7$. The transition cost is calculated by using the total of $\#(\mathcal{S}_{i} \rightarrow \mathcal{S}_{j})$ for all mental state. Eq.(\ref{eq:TransitionCost}) means that the higher transition cost is, the less transition occurs.

\begin{figure}[hbtp]
\begin{center}
\includegraphics[scale=0.38]{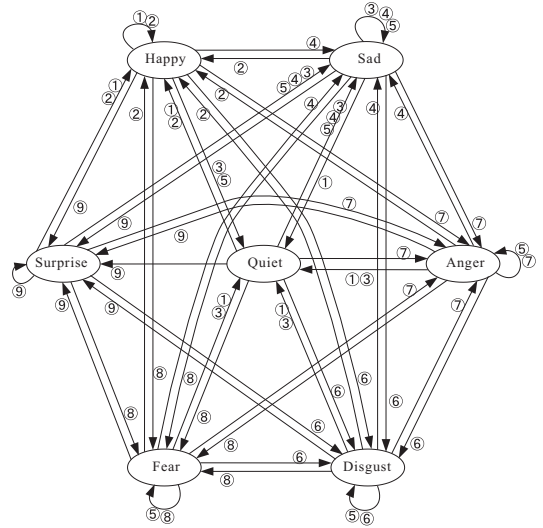}
\caption{MSTN with EGC}
\label{fig:MSTN_EGC}
\vspace{-5mm}
\end{center}
\end{figure}

Eq.(\ref{eq:TransitionCost_next}) calculates the next mental state from the current mental state $\mathcal{S}_{cur} \in \mathbf{S}$ by using the emotion vector.
\begin{equation}
next=\arg \max_{k} \frac{e_{k}}{cost(\mathcal{S}_{cur}, \mathcal{S}_{i})}, \ 1 \leq k \leq 9
\label{eq:TransitionCost_next}
\end{equation}
The emotion vector consists of 9 kinds of emotion groups which are classified 28 kinds of emotions as shown in Table \ref{tab:classofemotion}. Fig.\ref{fig:MSTN_EGC} shows the MSTN by using EGC. The circled numbers in Fig.\ref{fig:MSTN_EGC} are the number in the left side of Table \ref{tab:classofemotion}. The $e_{k}$ $(1 \leq k \leq 9)$ shows the strength of emotion group $k$ and takes the maximum value of elements belonged in each set $e_{k}$ as follows.

\begin{description}
\item $e_{1}=\max (e_{gloating}, e_{hope}, \cdots, e_{shy})$
\item $e_{2}=\max (e_{joy}, e_{happy\_for})$
\item $\vdots$
\item $e_{9}=\max (e_{surprise})$

\end{description}
\begin{table}[tbp]
\begin{center}
\caption{Classification of Generated Emotion}
\begin{tabular}{c|c}
\hline
No. & Emotion \\ \hline
  & gloating, hope, satisfaction, relief, pride,\\
1 & admiration, liking, gratitude, gratification,\\
  & love, shy \\ \hline
2 & joy,  happy\_for \\ \hline
3 & sorry-for, shame, remorse \\ \hline
4 & fear-confirmed, disappointment, sadness \\ \hline
5 & distress, perplexity \\ \hline
6 & disliking, hate \\ \hline
7 & resentment, reproach, anger \\ \hline
8 & fear \\ \hline
9 & surprise \\ \hline
\end{tabular}
\label{tab:classofemotion}
\end{center}
\end{table}

The $emo$ in Eq.(\ref{eq:selectemotion}) calculates the maximum emotion group according to the transition cost between current state and next state.
\begin{equation}
emo_{k}=\arg \max_{k} \frac{e_{k}}{cost(\mathcal{S}_{cur}, next(\mathcal{S}_{cur},k))},  \ 1 \leq k \leq 9, 
\label{eq:selectemotion}
\end{equation}
where $next(\mathcal{S}_{cur}, k)$ is next mental state from the current state by selecting emotion group $k$.

\section{Concierge System for Tourists}
\label{sec:AndroidConciergeSystem}
Located in Hiroshima, which is remote from big cities, it is well-known widely as a sightseeing spot of World Heritage Site and a lot of guests seem to come all the way from Kyoto, Osaka, and Tokyo. Certainly, Hiroshima has two spots of UNESCO World Heritage Site \cite{UNESCO}. However, the other sightseeing spots are not known to the tourists. Then, they don't know how to enjoy in Hiroshima.

We developed a concierge system for Tourists in Hiroshima, which can recommend the sightseeing spots near the current place while they can enjoy conversation about weather forecasts and other miscellaneous information arranged only to the user by EGC. This section explains two functions, dialogue system to analyze the user's current emotion and recommendation system for sightseeing spots.

\subsection{Dialogue System with Android EGC}
\begin{figure}[hbtp]
\begin{center}
\subfigure[Android EGC (Mika)]{
\includegraphics[scale=0.2]{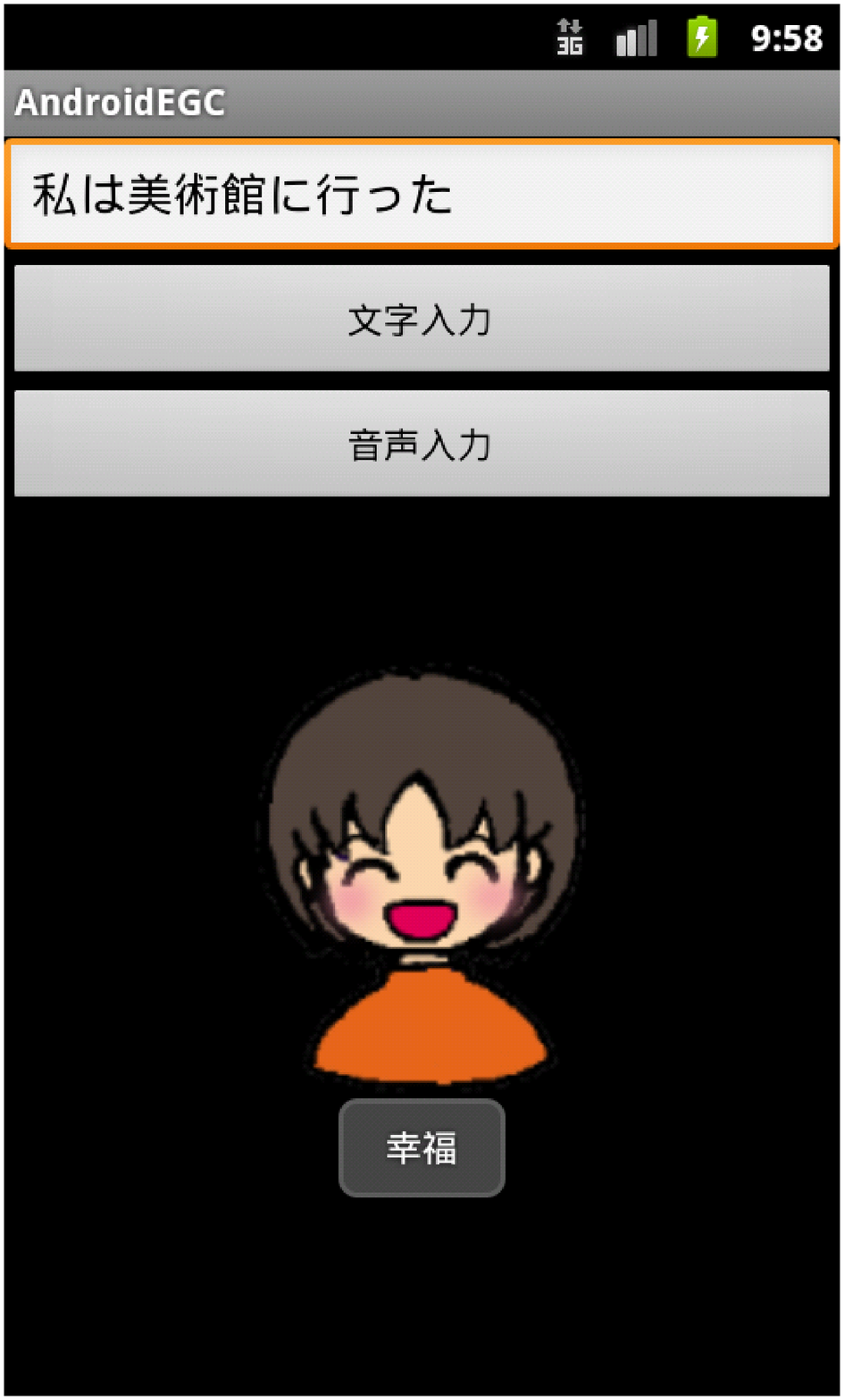}
\label{fig:AndroidEGC_Mika}
}
\subfigure[Android EGC (Takeshi)]{
\includegraphics[scale=0.2]{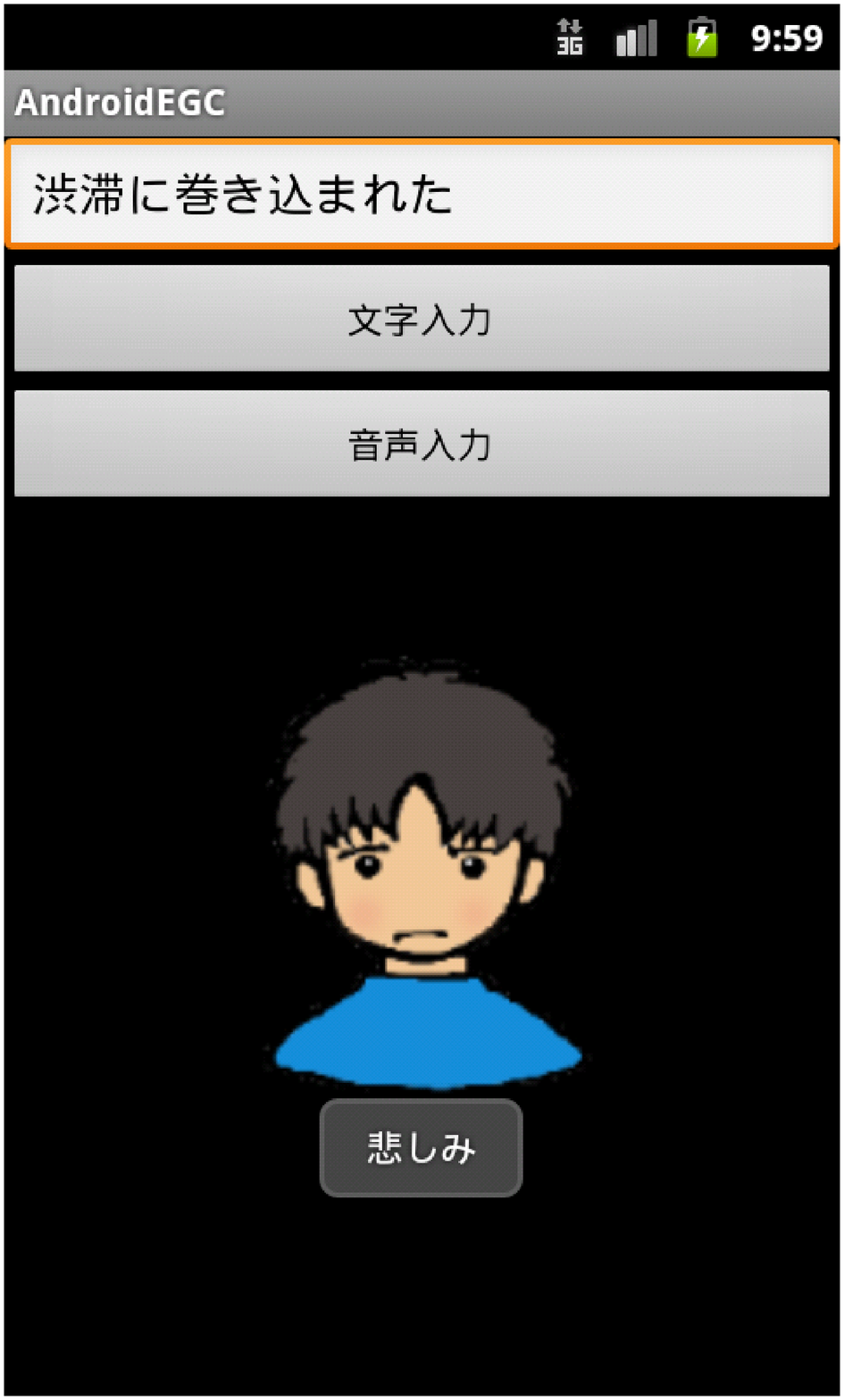}
\label{fig:AndroidEGC_Takeshi}
}
\caption{Dialogue by Android EGC}
\label{fig:AndroidEGC}
\end{center}
\end{figure}

In order to realize this system, we developed the Android EGC in Android smartphone\cite{AndroidEGC}, by embedding the MSTN and EGC. Because the EGC uses the PC based morphological analyzer and parser, the EGC in Android smartphone cannot operate well. Then, we developed the Android EGC by using the morphological analyzer in Yahoo API \cite{YahooAPI}. Android smartphone can use the voice recognition function in the Google Mobile App. Then our developed system recognizes the user's speech. Fig. \ref{fig:AndroidEGC} is the emotion type and its value calculated by the Android EGC. (The emotion value is not shown in the display.) The dialogue system displays the analysis result of the user emotion. Furthermore, we developed Android EGC API to embed in other applications.

\subsection{Recommendation System for Sightseeing Spots}
In our developed concierge system for tourists, the system can measure the current feeling of the user by Android EGC API. However, the spot image should be given to the system, because the tourist will have the imagery for each sightseeing spot. Then the questionnaire concerning the feelings co-occurring with sightseeing spot was filled out anonymously for 24 university students. 24 participants consists of 15 males and 9 female.

Miyajima and Atomic Dome are well known as sightseeing spots. However, the feeling to their spots are different. For example, Table \ref{tab:feeling_Miyajima} shows the average and the standard deviation of evaluation in five grades $\{0, 1, 2, 3, 4\}$ for user's feeling at sightseeing spot, Miyajima. On the contrary, the value of disgust at Atomic dome was high.

\begin{table}[!ht]
\begin{center}
\caption{Feelings to Miyajima}
\scalebox{0.8}[1.0]{
\begin{tabular}{|c|c|c|c|c|c|c|}
\hline
&happy&angry&surprise&sad&disgust&fear \\ \hline
average&{\bf 0.789}&0.039&0.421&0.079&0.039&0.079 \\ \hline
standard deviation&0.233&0.122&0.363&0.182&0.122&0.200 \\
\hline
\end{tabular}}
\label{tab:feeling_Miyajima}
\end{center}
\end{table}

Next questionnaire concerning the change of feeling before or after the occurrence of some events during the travel was investigated simultaneously. In the questionnaire, for example, the assumption of the situation is given as follows.
\begin{itemize}
\item You will visit the famous restaurant. How are you feeling?
\item However, the restaurant is closed. How are you feeling?
\end{itemize}
There are 19 questions concerning various situations during travels. The questions are created based on trouble situations in the travel website and the blog. Table \ref{tab:feelingchange} shows the average and the standard deviation of evaluation in five grades $\{0, 1, 2, 3, 4\}$ for the change of user's feeling among 24 participants. That is, these valuse suggest the user's average impression.

\begin{table}[!ht]
\begin{center}
\caption{The feeling changes}
\scalebox{0.7}[1.0]{
\begin{tabular}{|c|c|c|c|c|c|c|}
\multicolumn{7}{c}{Before the event}\\ \hline
&happy&angry&surprise&sad&disgust&fear \\ \hline
average&{\bf 3.625}&0.125&1.375&0.292&0.292&0.583 \\ \hline
standard deviation&0.484&0.331&1.184&0.538&0.5384&0.954\\ \hline
\multicolumn{7}{c}{After the event}\\ \hline
&happy&angry&surprise&sad&disgust&fear \\ \hline
average&0&2.75&{\bf 3.5}&3.375&1.958&0.458 \\ \hline
standard deviation &0&0.8292&0.5773&0.9492&1.1357&0.7626 \\
\hline
\end{tabular}}
\label{tab:feelingchange}
\end{center}
\end{table}

Our developed system can recommend 10 sightseeing spots in Hiroshima city and its suburb. The information related to sightseeing spots along the way from the current position to the recommended place is given. The system can measure the change of feelings through the dialogue during traveling.

\begin{figure}[hbtp]
\begin{center}
\subfigure[Recommended Spot list]{
\includegraphics[scale=0.2]{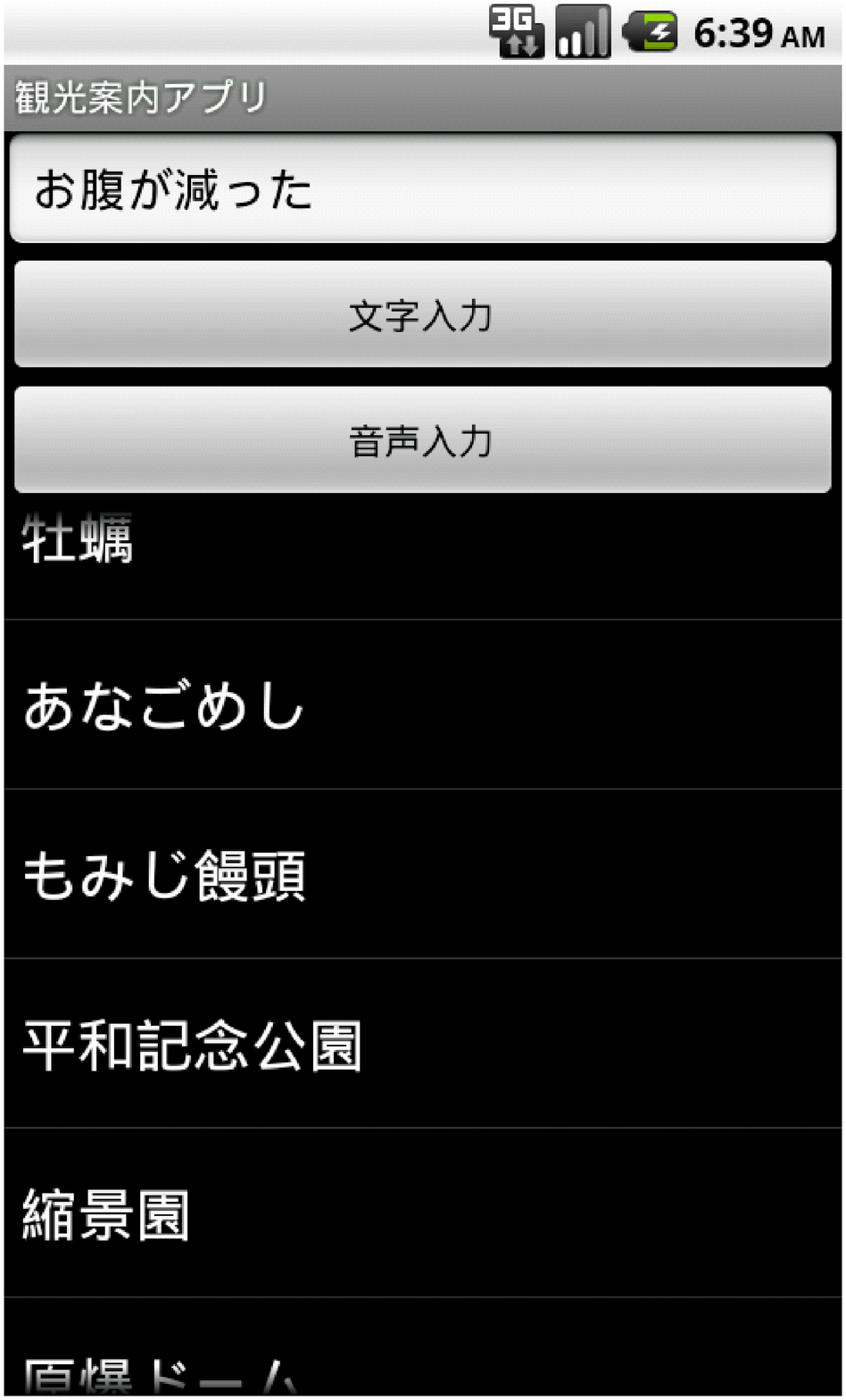}
\label{fig:RecommendList}
}
\subfigure[Recommended Information]{
\includegraphics[scale=0.2]{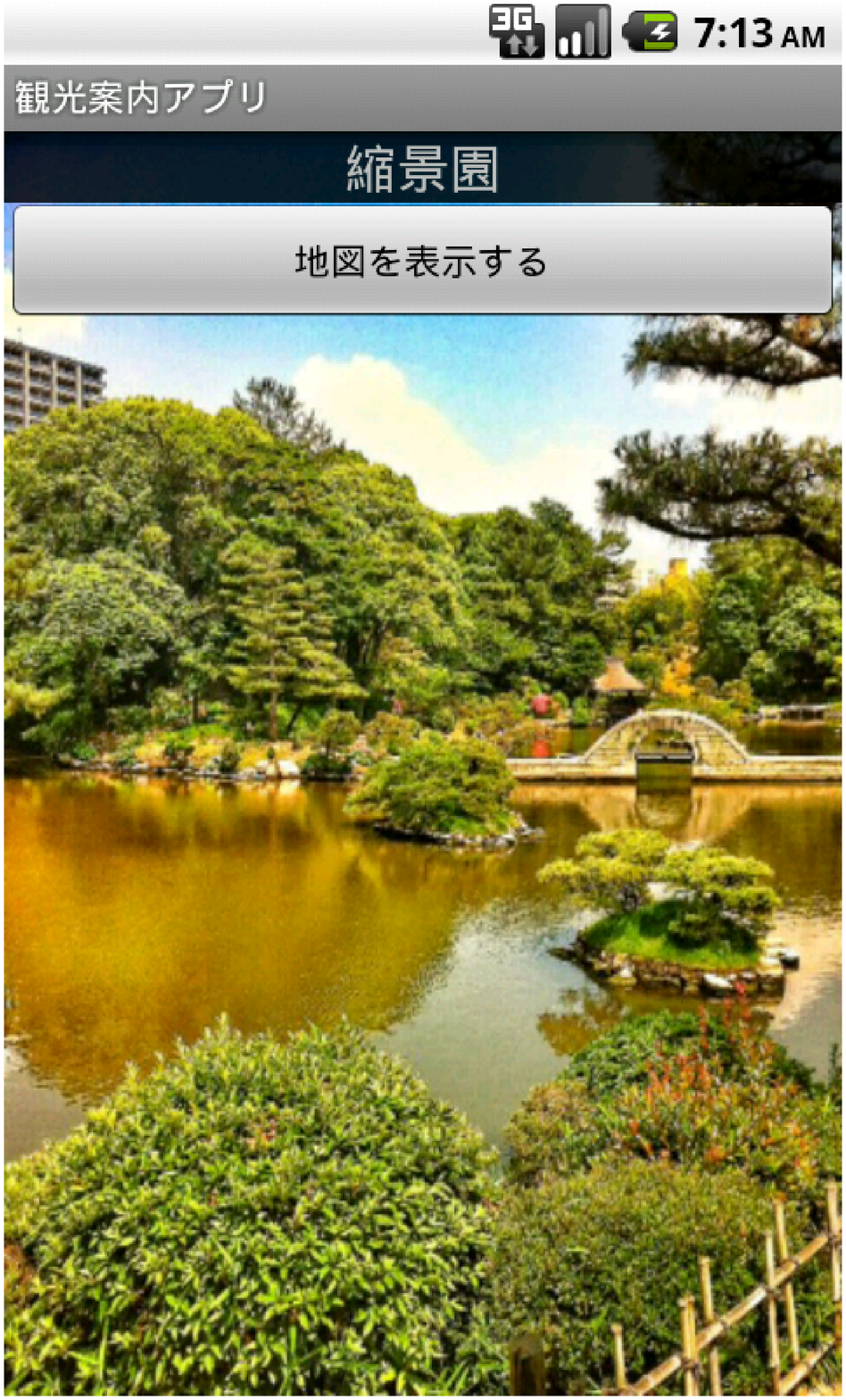}
\label{fig:RecommendSpot}
}
\caption{Recommendation System}
\label{fig:RecommendSystem}
\end{center}
\end{figure}

The system recommend the sightseeing spot which is the nearest emotion value as shown in Table \ref{tab:feeling_Miyajima}. Fig. \ref{fig:RecommendList} is the recommended list of sightseeing spots and Fig. \ref{fig:RecommendSpot} shows the information of the user selected spot.

\section{Conclusive Discussion}
\label{sec:Conclusivediscussion}
The smartphone can use various kinds of application such as web browser, e-mail, Google map and so on. Especially, the voice recognition function is the outstanding application to spread the capability of mobile phone, because the current dialogue system requires the user's typing. For example, the concierge system uses voice recognition function. However, the essential quality of dialogue is limited to question and answer, although the recognition rate becomes good. In order to enjoy real conversation, the system should evaluate the user's emotion. Our developed Android EGC can measure the user's feelings in the mental state transition network. We also developed the concierge system for tourists. Many tourists seem to come all the way from Kyoto, Osaka, and Tokyo to Hiroshima. However, most of them feels very regretful because of the few sightseeing spots in Hiroshima. In order to improve such a problem, Hiroshima Prefecture and the tourist association developed the smartphone application called `Hiroshima Quest' which enables travelers to plan enjoyable trip by bookmarking information of tourist sites, writing review and communicating with other users. Moreover, the Hiroshima tourist website replenishes the variety of information for foreigners. The Android application `Hiroshima Sightseeing map'\cite{Android_Market} has been developed to one of Mobile Phone based Participatory Sensing system, because not only tourism association but the local citizens should give the innovative and attractive information in sightseeing to visitors. We will embed the Android EGC into our developed `Hiroshima Sightseeing map' in near future. The usability for the developed system will be investigated.


\begin{thebibliography}{1}

\bibitem{Fujiwara10}
K.Fujiwara and I.Daibo, `The Function of Positive Affect in a communication content:Satisfaction with Conversation and hand movement', {\it The Japanese Journal of Research on Emotions}, Vol.17, No.3, pp.180-188 (2010) (Japanese)

\bibitem{Strongman96}
K.T.Strongman, {\it The Psychology of Emotion, Theories of Emotion in Perspective}, John Wiley \& Sons (1996)

\bibitem{Power07}
M.Power and T.Dalgleish, {\it Cognition and Emotion -From Order to Disorder}, Psychology Press (2007)

\bibitem{Kimura08}
M.Kimura, I.Daibo, and M.Yogo, `The Study of Emotional Contagion from the Perspective of Interpersonal Relationships', {\it Social Behavior and Personality}, Vol.36, No.1, pp.27-42 (2008)

\bibitem{Nagamachi10}
M.Nagamachi and A.M.Lokman, {\it Innovations of Kansei Engineering (Industrial Innovation)}, CRC Press (2010)

\bibitem{Matsunaga09}
M.Matsunaga et al., `Associations among positive mood, brain, and cardiovascular activities in an affectively positive situation', {\it Brain Research}, Vol.1263, pp.93-103 (2009)

\bibitem{Foster08}
P.S.Foster, V.Drago et al. `Emotion Influences on Spatial Attention', {\it Neuropsychology}, Vol.22, No.1, pp.127-135 (2008)

\bibitem{Park10}
J.Y.Park, B.M.Gu et al.`Integration of Cross-Modal Emotional Information in the Human Brain: an fMRI study, Cortex', {\it a journal devoted to the study of the nervous system and behavior}, Vol.46, No.2, pp.161-169 (2010)

\bibitem{Ichimura03}
T.Ichimura, T.Yamashita, K.Mera et al., `Emotion orientated intelligent systems', In {\it Internet-based Intelligent Information Processing Systems}, R.J.Howlett, N.S.Ichalkaranje, L.C.Jain, G Tonfoni Eds., pp.183-226, World Scientific Publishing Company (2003)

\bibitem{Mera03a}
K.Mera, 'Emotion Orientated Intelligent Interface', Doctoral Dissertation, Tokyo Metropolitan Institute of Technology, Graduate School of Engineering (2003)

\bibitem{Mera03b}
K.Mera, T.Ichimura, and T.Yamashita, `Complicated Emotion Allocating Method based on Emotion Eliciting Condition Theory', {\it Journal of the Biomedical Fuzzy Systems and Human Sciences}, Vol.19, No.1, pp.1-10 (2003)

\bibitem{Mera05}
K.Mera, and T.Ichimura, `Emotion Generating Method on Human - Computer Interfaces' In \emph{Computationally Intelligent Hybrid Systems: The Fusion of Soft Computing and Hard Computing}, Seppo J. Ovaska Eds., pp.277-312 Wiley-IEEE Press (2005)

\bibitem{Elliott92}
C.Elliott, `The Affective Reasoner: A process model of emotions in a multi-agent system', Ph.D thesis, Northwestern University, The Institute for the Learning Sciences, Technical Report No. 32 (1992)

\bibitem{Ren06}
F.Ren, `Recognizing Human Emotion baed on appearance information and Mental State Transition Network', {\it IPSJ SIG Technical Report}, pp. 43-48 (Japanese) (2006)

\bibitem{Mera10b}
K.Mera, T.Ichimura, Y.Kurosawa, and T.Takezawa, `Mood Calculating Method for Speech Interface Agent by using Emotion Generating Calculation Method and Mental State Transition Network', {\it Journal of Japan Society for Fuzzy Theory and Intelligent Informatics}, Vol.22, No.1, pp.10-24 (Japanese) (2010)

\bibitem{Nasukawa00}
T. Nasukawa et al., `Easy to Use Practical Freeware for Natural Language Processing', {\it IPSJ magazine}, Vol.41, No.11, pp.1201-1238 (Japanese) (2000)

\bibitem{Mera02}
K.Mera, T.Ichimura et al., `Invoking Emotions in a Dialog System based on Word-Impressions', 
{\it Journal of Japan Society of Artificial Intelligence}, Vol.17, No.3, pp.186-195 (Japanese) (2002)

\bibitem{Ortony88}
A.Ortony, G.L.Clore, and A.Collins, {\it The Cognitive Structure of Emotions}, Cambridge University Press (1988)

\bibitem{UNESCO}
UNESCO World Heritage Centre
`Official launch of the World Heritage Convention 40th anniversary year',
{\it http://whc.unesco.org/} (2012)
Retrieved 2012-12-29.

\bibitem{AndroidEGC}
K.Tanabe, I.Tachibana, T.Ichimura,
`Development of Emotion Orientated Interface on Android Smartphone',
Proc. of FSS2012(to appear in 2012) 

\bibitem{YahooAPI}
Yahoo JAPAN, Japanese morphological analyzer, {\it http://developer.yahoo.co.jp/webapi/jlp/ma/v1/parse.html}, (2012), Retrieved 2012-06-24.

\bibitem{Android_Market}
ITProducts `Hiroshima Tourist Map',
{\it https://market.android.com/details? id=jp.itproducts.KankouMap},
Retrieved 2011-11-15. (2011)

\end{thebibliography}
\end{document}